# Field-free superconducting diode effect of NbSe$_2$ induced by strain

Jiajun Li[1,3#], Minhao Zou[1,3#], Fengyi Guo[1,3#], Dai Zheng[1,3], Yiying Zhang[1,3], Yu Du[1,3,6] Fuwei Zhou[1,3], Heng Zhang[1,3], Wuyi Qi[1,3], Tianqi Wang[1,3], YeFan Yu[1,3], Rui Wang[1,3], Fucong Fei[1,4*], Hao Geng[2*], Fengqi Song[1,3,5*]

[1] *National Laboratory of Solid State Microstructures, Collaborative Innovation Center of Advanced Microstructures, Nanjing University, Nanjing 210093, China.*

[2] *College of Physics, Nanjing University of Aeronautics and Astronautics, Nanjing 211106, China.*

[3] *School of Physics, Nanjing University, Nanjing 210093, China.*

[4] *School of Materials Science and Intelligent Engineering, Nanjing University, Suzhou 215163, China.*

[5] *Institute of Atom Manufacturing, Nanjing University, Suzhou 215163, China.*

[6] *Suzhou Polytechnic University, Suzhou 215104, China.*

*Corresponding authors: F. Fei (feifucong@nju.edu.cn); H. Geng (genghao@nuaa.edu.cn); F. Song (songfengqi@nju.edu.cn).

**Abstract**

Superconducting diodes, similar to semiconductor diodes, possess unidirectional superconducting properties and are the fundamental units for constructing superconducting quantum computing, thus attracting widespread attention. At present, most of superconducting diodes require an external magnetic field or proximity effect to break time reversal symmetry (TRS). The cases of intrinsic superconducting diode effect (SDE) under zero magnetic field are relatively scarce, and there are still some puzzles especially regarding the reasons for the TRS breaking. Here, we not only report field free SDE in NbSe$_2$ induced by strain, but also large values of the difference of $I_{c+}$ and $|I_{c-}|$ ($\Delta I_c$) of 286 μA and the superconducting diode efficiency ($\eta$) of 6.76 % are achieved. Interestingly, $\Delta I_c$ varies with the magnetic field and exhibits two distinct evolutionary behaviors with B-odd or B-even symmetry in various devices. We attribute this to the selective activation of two independent, spatially-orthogonal mechanisms: a stress-induced real-space polarity and a field-induced reciprocal-space asymmetric energy bands. In general, we propose an extremely effectively method to produce field free SDE, even when the material itself does not possess field free SDE, and provide new perspectives to understand the SDE which build new avenues for superconducting quantum devices.

**Introduction**

Semiconductor diodes, due to their unidirectional conductivity, form the basic unit of modern electronic devices. Correspondingly, the diode effect that occurs in superconductors is called the superconducting diode effect (SDE). That is, the critical current in one direction (positive) is not equal to the critical current in the opposite (negative) direction. When a single-directional current with an amplitude between the two different critical currents is applied, the superconducting diode remains in a zero-resistance state in one direction and exits the superconducting state when the current direction changes[1, 2]. Compared to semiconductor diodes, superconducting diodes have received widespread attention due to their advantages such as reduced energy loss, higher current density and greater compatibility with quantum systems. Therefore, superconducting diodes play an important role in quantum computing, quantum information and communication technology[3, 4].

The superconducting diode effect (SDE) needs inversion symmetry and time reversal symmetry (TRS) breaking[5-7]. Usually, non-centrosymmetric superconductors and artificially constructed heterojunctions with asymmetric structures are utilized to break the inversion symmetry[8-12]. For TRS, it is usually broken by applying an external magnetic field or proximity effect [3, 12-17], like Nb/V/Ta superlattice [18], $NbSe_2/Fe_3GeTe_2$[12, 13], EuS/V[17] and so on. Furthermore, in specific systems, due to certain factors such as valley polarization[19, 20], interface magnetism[21] and magnetic moment of the orbit [22], this can cause the system to exhibit TRS breaking and then zero-field SDE is able to be observed. This provides further possibilities for future superconducting quantum computing. However, there are still many unanswered questions regarding the mechanism of zero-field SDE. Some systems such as $PbTaSe_2$[23] and $Bi_2Sr_2CaCu_2O_{8+\delta}$ (BSCCO)[24] exhibit the characteristic that the difference of the forward and reverse superconducting critical currents ($\Delta I_c$) have even symmetry with respect to the magnetic field and the diode effect is most pronounced at zero magnetic field, which cannot be fully explained by the spontaneous breaking of time-reversal symmetry in the system itself.

In this study, we report the observation of field free SDE in $NbSe_2$ induced by strain. $NbSe_2$ belongs to the hexagonal space group *P6₃/mmc,* and is a centrosymmetric superconductor, while few odd-layered $NbSe_2$ is a non-centrosymmetric superconductor[25]. By controlling the curvature of the substrate, local stress is applied in certain areas on the same thin-layer NbSe$_2$ device. Electrical transport experiments reveal that significant SDE appears in the stressed regions, while no such effect can be achieved in the non-stressed areas. Meanwhile, in the stress regions of different devices, two different kinds of symmetry of SDE are amazingly observed, whose $\Delta I_c$ are symmetric and antisymmetric with respect to the magnetic field respectively. Similar phenomena was observed in previous studies on PbTaSe$_2$[23]. However, in comparison, the value of $\Delta I_c$ and the superconducting diode efficiency ($\eta$) in stressed NbSe$_2$ have been enhanced by several orders of magnitude, reaching 286 μA and 6.76% under zero field respectively. These values are quite large for devices fabricated by traditional superconductors without junctions or domains, and are only smaller than the SDE report on high-temperature superconductor BSCCO [19, 22-24]. To explain these phenomena, we introduce a theoretical interpretation based on two independent and spatially-orthogonal non-reciprocal mechanisms, which can also be used to explain previous studies[23, 24]. The first mechanism, analogous to the non-reciprocal transport and Josephson diode mechanisms that arise solely from broken space-inversion symmetry[29, 30], is activated by uniaxial stress to produce the B-even SDE along the armchair direction. The second, a finite-momentum for Cooper pairs mechanism requiring the breaking of both TRS and space-inversion symmetry[5-7], is switched on by the magnetic field to produce the B-odd SDE along the zigzag direction. Our work provides a robust experimental platform for producing and studying zero-field SDEs, especially for understanding the generation mechanism of SDEs with magnetic field symmetry. It also offers potential application possibilities for future superconducting quantum computing.

**Results and discussion**

2H-NbSe$_2$ is a type-II Ising superconductor with a superconducting temperature around 7 K[26]. The Nb atoms are arranged in a trigonal prismatic coordination environment, surrounded by six Se atoms. The 2H-phase is composed of ABAB stacking arrangement, and there are only van der Waals forces between the layers. Therefore, it is very conducive to mechanical exfoliation. Few odd-layered NbSe$_2$ is non-centrosymmetric superconductor with inversion symmetry breaking. In order to introduce stress into NbSe$_2$, we vaporized aluminum and deposited an aluminum nanoridge onto the silicon wafer by thermal evaporation, and then melt the aluminum nanoridge to forms droplets with curved surface. After cooling, the surface of aluminum remains curved. Subsequently, polydimethylsiloxane (PDMS) is used to transfer insulating spacer of hexagonal boron nitride (h-BN) and NbSe$_2$ flake onto the aluminum in turn. Under the support of Al, the curved aluminum surface applied a uniaxial stress to NbSe$_2$. The underlying Al nanoridge exert a supporting force on the NbSe$_2$ on top, causing the NbSe$_2$ nanosheets on the top Al nanoridge to bend, introducing uniaxial stress in NbSe$_2$ flake. The schematic diagram of the device is shown in Figure 1(a), and the corresponding optical photograph of a typical device is displayed in the inset of Figure 1(b). The specific processing procedure can be found in the **Method** section. The temperature-dependent resistance of Device 1# (strain area) is shown in Figure 1(b). We take the temperature corresponding to 90% of the longitudinal resistance as the superconducting phase transition temperature, and the superconducting transition temperature ($T_c$) of device 1# is 6.4 K. Figure 1(c) and (d) respectively show the temperature-dependent resistance under different magnetic fields when temperatures ranging from 2 K to 10 K, as well as the magnetic field-dependent curves at different temperatures. As the magnetic field increases, the superconductivity of NbSe$_2$ is gradually suppressed. Meanwhile, as the temperature decreases, the magnetic field required to suppress superconductivity becomes greater, which is consistent to the typical superconducting behavior of NbSe$_2$[27, 28]. Subsequently, In order to study the effect of stress on NbSe$_2$ nanosheets, we measure the *I-V* curves with sweeping the current from 0 to positive and from 0 to negative values, and the results are shown in Figure 1(e). Interestingly, under zero-field conditions, the forward critical current $I_{c+}$ is apparently larger than the reverse critical current $|I_{c-}|$, indicating the existence of field-free superconducting diode effect. To further verify that the superconducting diode effect is caused by stress, we conduct measurements in the non-stressed area of the same device. The results are shown in Figure 1(f) and no superconducting diode effect is observed.

In order to investigate the behavior of field-free SDE observed above in detail, we further measure the *I-V* curves of Device 1# under different external magnetic fields and various temperatures. The evolution behavior of the SDE under various magnetic fields is shown in Figure 2(a). We extract the value of critical currents $I_{c+}$($|I_{c-}|$) from the *I-V* curve in positive(negative) current region under different magnetic fields and plot the graph as shown in Figure 2(b). One can see that the value of $I_{c+}$ and $|I_{c-}|$ are symmetric with respect to the magnetic field $B = 0$, and gradually decreases as the magnetic field increases. We further calculate the difference between the forward and reverse critical currents($\Delta I_c = I_{c+} - |I_{c-}|$ ) and calculate the superconducting diode efficiency $\eta$ with the formula $\eta = \frac{I_+ - I_-}{I_+ + I_-} \times 100\%$. The results are shown in Figure 2(c). It is clear that $\Delta I_c$ is also symmetric with respect to the magnetic field $B$ = 0 Oe. At zero field, the maximum value of $\Delta I_c$ = 286 μA is achieved, and at the same time, the superconducting diode efficiency reaches its maximum of 6.76% which has high efficiency in traditional superconductors without junctions and magnetic field[19, 22-24]. As the magnetic field increases, the value of $\Delta I_c$ and $\eta$ begin to decrease rapidly near the low magnetic field. When the magnetic field is large than 1000 Oe, the $\Delta I_c$ and $\eta$ gradually decrease to zero. Figure 2(d) and (e) respectively show the variations of the critical currents, as well as $\Delta I_c$ and $\eta$ under various temperatures. As the temperature increases from 2 K to 5.8 K, the superconductivity is gradually suppressed, resulting in a gradual reduction of the critical current. Correspondingly, $\Delta I_c$ and $\eta$ also gradually decrease, and become 0 at 5.8 K.

Furthermore, to investigate the SDE effect in more devices, transport measurements on various device with similar structures as Figure 1(a) are carried out. A kind of SDE behavior that is different from the one in device 1# is unveiled, which SDE has B-even symmetry as seen in Figure 3(a)-(f). In Figure 3(a), as the magnetic field changes from negative to positive, the polarity reversal of the superconducting diode of device 2# occurs. By extracting $I_{c+}$ and $|I_{c-}|$ under different magnetic fields, the variation of critical current with magnetic field can be plotted as seen in Figure 3(b). The values of $I_{c+}$ and $|I_{c-}|$ exhibit a similar decreasing trend comparing with device 1# with increasing magnetic field. However, the peak values of $I_{c+}$ and $|I_{c-}|$ do not hold at $B$ = 0 Oe, but deviate respectively towards the positive and negative magnetic fields. Thus, in figure 3(d), $\Delta I_c$ and $\eta$ of device 2# show an antisymmetric behavior with respect to the magnetic field which is B-odd symmetric that is different from B-even symmetry in device 1#. Similar phenomena are also observed in device 3, as shown in figure 3(b). Figures 3b and e, similar to figures 3a and c, all show a reversal of diode polarity. However, the SDE under negative fields is more pronounced compared to positive fields, and $\Delta I_c$ is more significantly. The value of $\Delta I_c$ and $\eta$ versus magnetic field are shown in figure 3(f), and unlike device 2# which is strictly B-odd symmetric, device 3# has a certain offset and is not strictly B-odd symmetric. Instead, it contains a component similar to the B-even symmetry of device 1, existing in mixing of B-even and B-odd symmetry. Therefore, device 3# can also observe a considerable SDE in the absence of an external field. As the magnetic field increases, the superconductivity of the sample is suppressed, and $I_{c+}$ and $|I_{c-}|$ gradually decrease. Correspondingly, $\Delta I_c$ and $\eta$ also gradually decrease to values close to 0. From the measurement results of these devices, it can be seen that the SDE caused by pressure, which varies with the magnetic field, contains three situations. One is B-even symmetry, one is B-odd symmetry, and the last one is a mixed case of B-even and B-odd. We speculate that this is because the current flows along or perpendicular to the direction of lattice distortion.

To explain the above experimental phenomena from a theoretical perspective, we further discuss the mechanism of generating SDE in NbSe$_2$. The physical origin of the superconducting diode effect (SDE) is rooted in the breaking of system symmetries. The core of our proposed explanation is the existence of two SDE mechanisms in NbSe$_2$ that are physically independent and spatially orthogonal. The first is a polarization mechanism originating from the breaking of real-space lattice symmetry, characterized by an axis $m_p$; the second is a valley-polarization mechanism originating from the asymmetry of the reciprocal-space band structure, characterized by an axis $m_v$. For a hexagonal lattice, these two axes are precisely perpendicular, as illustrated in Figure 4. Crucially, the activation conditions for these two mechanisms are independent: uniaxial stress is the necessary condition to activate the real-space polarization SDE, whereas an external magnetic field is necessary to activate the reciprocal-space valley-polarization SDE. Therefore, the ultimate manifestation of the SDE—including its zero-field behavior and its even/odd parity with respect to the magnetic field—depends entirely on the experimental conditions (the presence of stress and/or a magnetic field) and which activated axis the transport current is parallel to.

When the current flows along the armchair direction, the current direction is in parallel with the real-space polarization axis $m_p$ (see Figure 4(a) and (b)). According to our experimental observations, the zero-field SDE appears only after applying stress, indicating that stress is the exclusive switch for activating this channel. The fixed lattice distortion caused by stress endows the superconductor with an intrinsic structural polarity, making its behavior analogous to a "p-n junction"[29, 30]. This breaking of space-inversion symmetry, "frozen" into the lattice by stress, is sufficient to directly generate an SDE under zero-magnetic-field ($B_z = 0$) conditions. In this configuration, an external magnetic field acts merely as a secondary perturbation, modulating the pre-existing, structurally determined non-reciprocity without reversing its direction. Consequently, the degree of non-reciprocity is insensitive to the direction of the magnetic field $B_z$, and its magnitude exhibits a symmetric, even-function dependence on the field, i.e., $\Delta I_c(B_z) \approx \Delta I_c(-B_z)$.

When the current flows along the zigzag direction, which parallels with the reciprocal-space valley symmetry axis $m_v$ (see Figure 4(c) and (d)), the SDE in this channel is governed by the valley-polarization and finite-momentum pairing mechanism [5, 6, 31]. In this case, the external magnetic field plays essential role to activate the SDE. At zero magnetic field, time-reversal symmetry is preserved, and this channel remains "off", producing no SDE. When an out-of-plane magnetic field $B_z$ is applied, the Zeeman effect acts as the driving force, breaking time-reversal symmetry and inducing valley polarization, which in turn causes the Cooper pairs to acquire a net momentum $q_0$ along the $m_v$ axis. This field-induced momentum $q_0$ in the equilibrium is known to drive the SDE [5]. Moreover, since $q_0$ is linearly proportional to the external field $B_z$, reversing the field ($B_z \rightarrow -B_z$) also reverses the momentum ($q_0 \rightarrow -q_0$), causing a complete reversal of the non-reciprocal direction of the SDE. This finally leads to the SDE exhibiting odd function dependence on the magnetic field, i.e., $\Delta I_c(B_z) = -\Delta I_c(-B_z)$.

**Conclusion**

Field free SDE in NbSe$_2$ devices is observed under stressed regions, while no SDE can be observed in the stress-free regions, indicating that field free SDE induced by stress. Furthermore, under external magnetic field, the SDE in strained NbSe$_2$ display two distinct behaviors with $\Delta I_c$ follows B-odd or B-even symmetry to the magnetic fields in different devices. We attribute the anisotropic superconducting diode effect to the selective activation of two independent and orthogonal non-reciprocal channels. A stress-induced real-space polarity governs the B-even SDE along the armchair direction, whereas a field-activated valley polarization mechanism dictates the B-odd SDE along the zigzag direction. This study has proposed an effective method for generating field-free SDE, which is expected to be applied to various superconducting materials. It also provides a favorable platform for researching the influence of stress on superconducting materials and a new approach for the establishment of superconducting quantum devices.

**Methods**

Aluminum nanoridge are fabricated through electron beam exposure and sputtered 45nm thick Au on silicon substrates. Heating the aluminum nanoridge to 680°C in a vacuum environment to melt it. Melting aluminum forms a droplet, and after cooling, the surface would keep a curved shape. A h-BN thin flake is first identified by optical microscopy and then transferred onto Aluminum nanoridge as an insulating spacer. $NbSe_2$ thin flake is mechanically exfoliated from $NbSe_2$ bulk single crystals on PDMS stamp and then is transferred onto h-BN. Gold electrodes with thickness of 60 nm are patterned on $NbSe_2$ thin flakes via electron-beam lithography and sputtering. The transfer processes are carried out in a glove box. The electrical properties are measured with a Physical Properties Measurement System (PPMS, Quantum Design) equipped with a Lock-in Amplifier (Stanford Research Systems, SR830) and a sourcemeter (Keithley 2400).

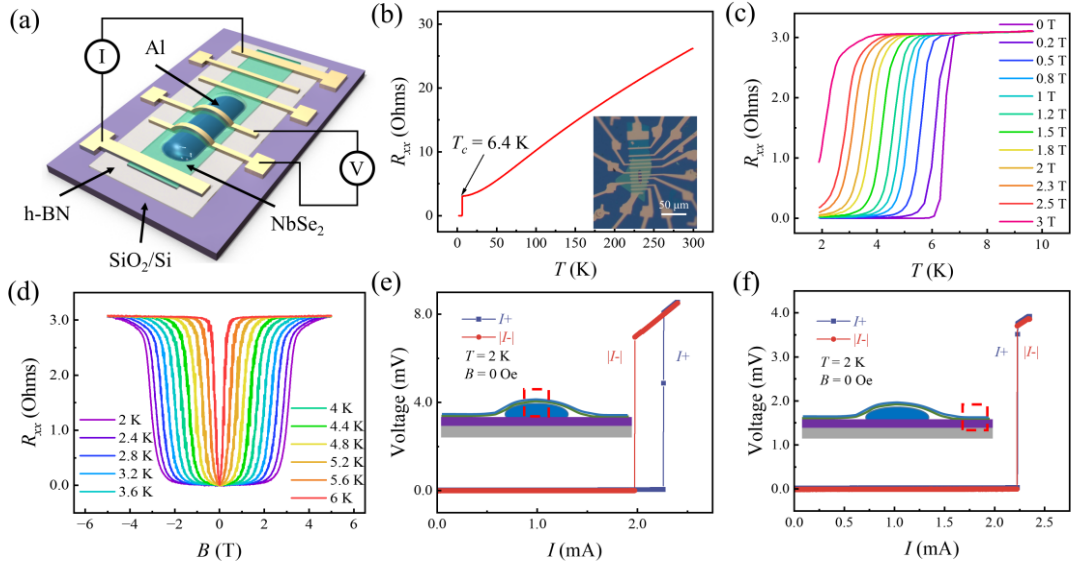

**FIG 1:** Field-free superconducting diode effect induced by strain. (a) Schematic of the NbSe$_2$ device with the Al nanoribbons to induce strain; (b) temperature dependence of the resistance, the inset shows the optical image of the NbSe$_2$ device with a scale bar of 5 μm, and the superconducting transition temperature $T_c$ = 6.4 K; (c) enlarged view of the cooling curve from 10 K to 2 K with different magnetic field from 0 T to 3 T; (d) magnetic field dependence of the resistance with different temperature from 2 K to 6 K; (e) and (f) are the *I-V* curve of stressed region and zero stressed region, showing the superconducting diode effect induced by strain. The blue and red lines respectively represent the forward and reverse currents. The red dotted lines in the illustration represent the electrodes used for measurement. In all measurement, *B // c*.

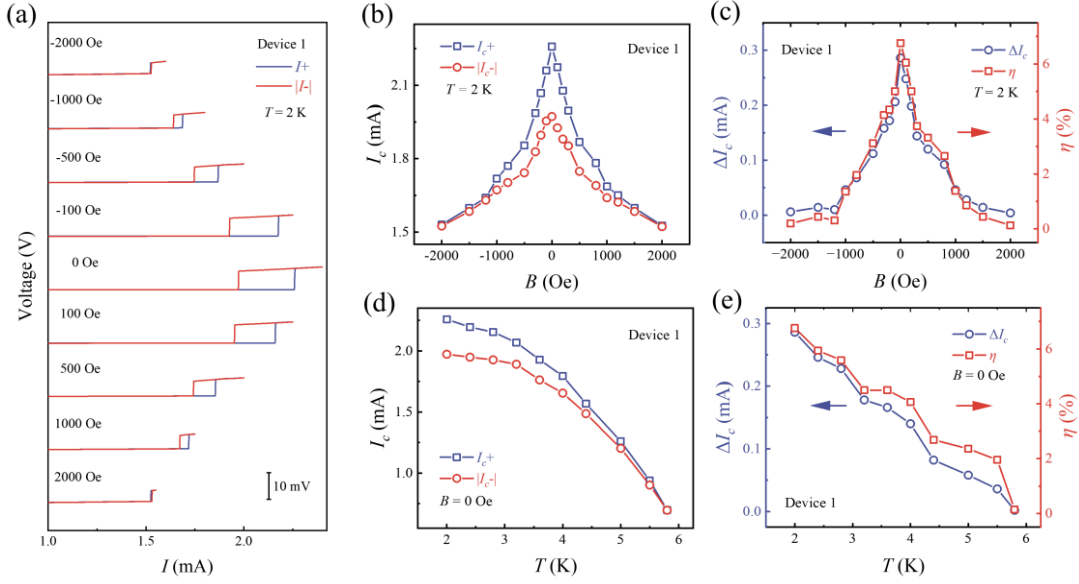

**FIG 2:** Magnetic field and temperature dependent *I-V* curves in NbSe$_2$ device 1. (a) the *I-V* curves in different magnetic field with $T = 2$ K; (b) magnetic field dependence of critical current with $T = 2$ K, the black and red dots respectively represent the forward and reverse critical currents; (c) The blue curve represents the curve showing the difference between the forward and reverse critical currents as a function of the magnetic field, and it is symmetrical around $B = 0$ Oe. Red curve represents superconducting diode efficiency under different magnetic field. (d) temperature dependence of forward and reverse critical current with $B = 0$ Oe; (e) plot of the difference between the forward and reverse critical currents and superconducting diode efficiency as a function of temperature. In all measurement, $B // c$.

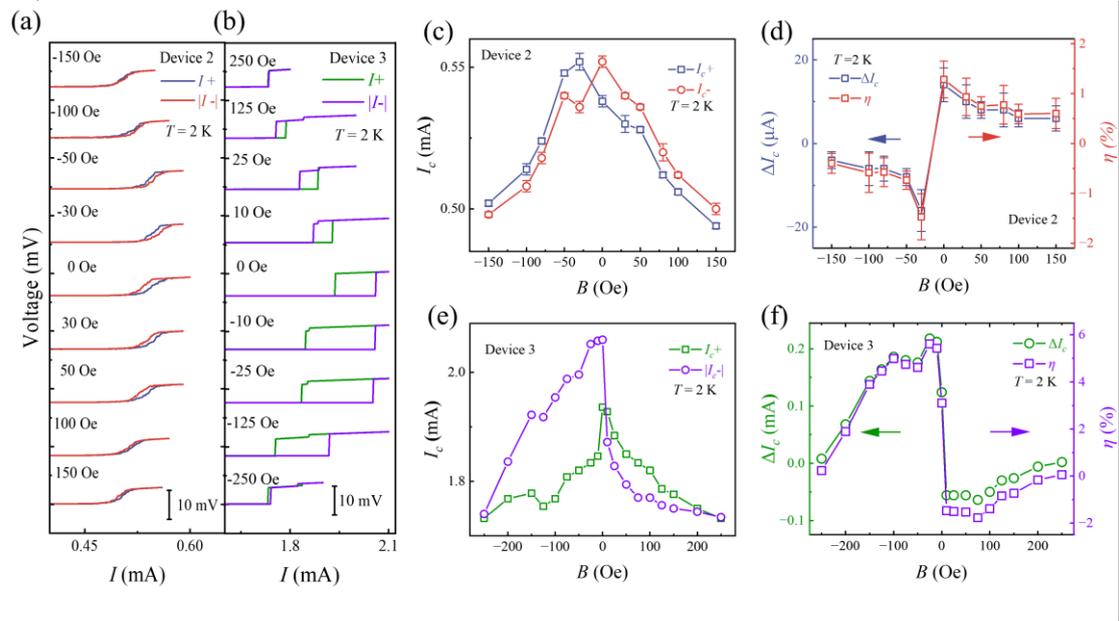

**FIG 3:** Magnetic field dependent *I-V* curves in NbSe$_2$ device 2 and device 3. (a) and (b) magnetic field dependence of the *I-V* curves of device 2 and device 3 with $T = 2$ K; (c) and (e) are the plot of the forward (green/blue) and reverse (purple/red) critical currents of device 2 and device 3 as the function of the magnetic field; (d) and (f) are magnetic field dependent difference between forward and reverse critical current ($\Delta I_c$) and superconducting diode efficiency ($\eta$) of device 2 and device 3 at $T = 2$ K. In all measurement, $B // $ c.

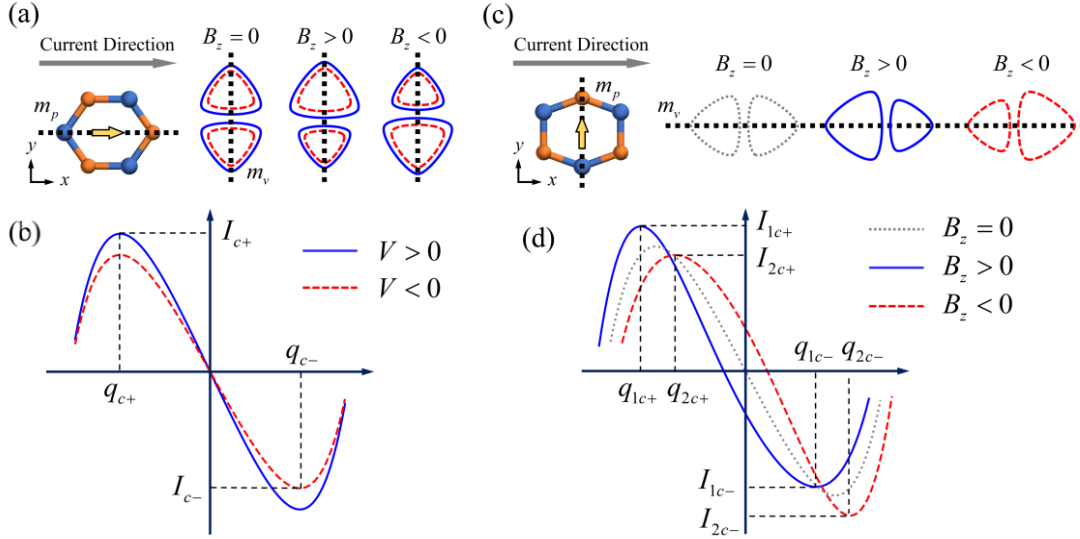

**FIG 4:** Direction-dependent SDE in strained NbSe$_2$ activated by independent, orthogonal mechanisms. (a) and (b) Armchair Direction (B-even SDE) - Stress-activated real-space mechanism. The current $I$ is parallel to the real-space polarization axis $m_p$. This SDE is activated by uniaxial stress and exists at zero field, exhibiting an even-function dependence on the external magnetic field. (b) The schematic $I$-$q$ plot illustrates its intrinsic asymmetry (solid blue and dashed red curves represent the case with opposite current direction with external voltage $V > 0$ and $V < 0$, respectively). (c) and (d) Zigzag Direction (B-odd SDE) - Field-activated reciprocal-space mechanism. The current $I$ is parallel to the reciprocal-space valley symmetry axis $m_v$. This SDE is activated by an external magnetic field and vanishes at zero field. The Fermi surface sketch at (c) illustrates how the magnetic field $B_z$ leads to valley polarization. (d) The $I$-$q$ plot below shows that the SDE's asymmetry reverses with the sign of the magnetic field $B_z$ (solid blue curve for $B_z > 0$, dashed red for $B_z < 0$), manifesting its odd-function character. The axes corresponding to the two mechanisms, $m_p$ and $m_v$, are mutually perpendicular.